\documentclass[aps,amsmath,amssymb,groupedaddress]{revtex4}
\usepackage{graphicx}

\newcommand{\nc}{\newcommand}
\nc{\beq}{\begin{equation}}
\nc{\eeq}{\end{equation}}
\nc{\beqa}{\begin{eqnarray}}
\nc{\eeqa}{\end{eqnarray}}

\def\gsim{\mathrel{\rlap{\lower4pt\hbox{\hskip1pt$\sim$}}
    \raise1pt\hbox{$>$}}}       %greater than or approx. symbol

\begin{document}

\title{On the origin of probability in quantum mechanics}

\author{Stephen~D.~H.~Hsu} \email{hsu@uoregon.edu}
\affiliation{Institute of Theoretical Science \\ University of Oregon,
Eugene, OR 97403}

\begin{abstract}
I give a brief introduction to many worlds or ``no wavefunction collapse'' quantum mechanics, suitable for non-specialists. I then discuss the origin of probability in such formulations, distinguishing between objective and subjective notions of probability.  
\end{abstract}

%\pacs{}

\maketitle

\date{today}

\bigskip

\section{Do wave functions collapse?}

Quantum mechanics, as conventionally formulated, has two types of time evolution. An isolated system $S$ evolves according to the Schrodinger equation
\begin{equation}
\label{unitary}
\vert \Psi (t) \rangle = \exp ( -i H t )~ \vert \Psi (0) \rangle ~,
\end{equation}
where the time evolution operator $U(t) = \exp ( -i H t )$ is unitary: $U^\dagger U = 1$. However, when a measurement is made the state undergoes non-unitary von Neumann projection (i.e., collapses) to an eigenstate corresponding to the eigenvalue observed by the measuring device. Because the two types of time evolution are so radically different, students typically demand (and indeed are entitled to demand) a rigorous definition of exactly when each of them apply. Unfortunately, as is widely acknowledged, the conventional interpretation does not supply an entirely satisfactory definition (see, e.g., {\it Against Measurement} by J.S. Bell \cite{Bell}, and the discussion of decoherence below).

Many students over the years (including the author, when he was a student) have wondered whether the measurement process can be described within a larger isolated system $S'$, containing both the original system $S$ and the measuring apparatus $M$, whose dynamics is still unitary. However, measurement collapse and unitary evolution of $S'$ are incompatible, because the projection operation is not invertible. Collapse is fundamentally irreversible, whereas unitary evolution of the system and measuring device together is reversible. Nevertheless, as we shall discuss below, unitary evolution of the larger system $S'$ is compatible with the {\it appearance} of collapse to observers within the system.

To make this discussion more explicit, let $S$ be a single qubit and $M$ a device (e.g., Stern-Gerlach device) which measures the spin of the qubit along a particular axis. The eigenstates of spin along this axis are denoted $\vert \pm \rangle$. We define the operation of $M$ as follows
\begin{eqnarray}
\vert + \rangle \otimes \vert M \rangle ~\longrightarrow ~  \vert + \, , \, M_+ \rangle  \nonumber \\
\vert - \rangle \otimes \vert M \rangle ~\longrightarrow ~  \vert - \,,\, M_- \rangle 
\end{eqnarray}
where $M_+$ denotes a state of the apparatus which has recorded a $+$ outcome, and similarly with $M_-$. We can then ask what happens to a superposition state $\vert \Psi_S \rangle =  c_+ \vert + \rangle ~+~ c_- \vert - \rangle$ which enters the device $M$. In the conventional formulation, with measurement collapse, {\it one} of the two final states $\vert + \, , \, M_+ \rangle~$ {\it or} $~\vert - \, , \, M_- \rangle$ results, with probabilities $\vert c_+ \vert^2$ and $\vert c_- \vert^2$ respectively. This probability rule is called the Born rule. In the conventional formulation the notion of probability enters precisely when one introduces wave function collapse. Without collapse the status of probability is much more subtle, as we will see.

\begin{figure}
\includegraphics[width=16cm]{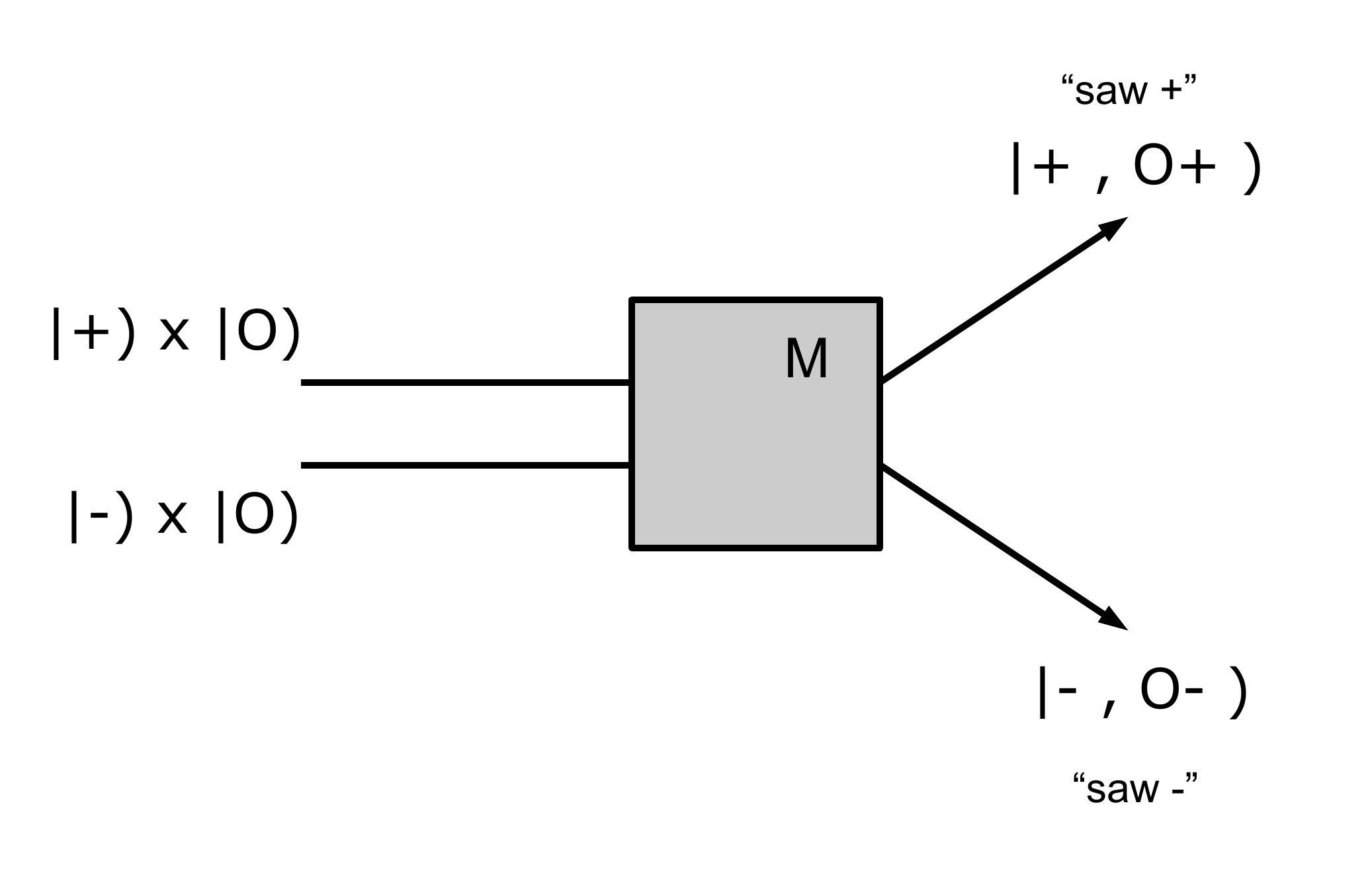}
\caption{An observer $\cal O$ uses a device $M$ to measure a spin state $\vert \pm \rangle$. Before the measurement there is a single observer state $\vert {\cal O} \rangle$, but afterwards there are two observer states $\vert {\cal O}_\pm \rangle$.}
\label{identity1}
\end{figure}

However, if the combined system $S' = S + M$ evolves unitarily (in particular, {\it linearly}) as in (\ref{unitary}), we obtain a superposition of measurement device states (see figure 1):
\begin{equation}
\big(  c_+ \vert + \rangle ~+~ c_- \vert - \rangle  \big ) \otimes \vert M \rangle ~~ \longrightarrow ~~
c_+ \, \vert + \, , \, M_+ \rangle ~+~ c_- \, \vert - \, , \, M_- \rangle ~~.
\end{equation}
At first, this seems counter to actual experience: properly executed measurements produce a single outcome, not a superposition state. But how do we know for sure? In fact, an observer described by the state $M_+$ might be unaware of the second branch of the wave function in state $M_-$ for dynamical reasons related to a phenomenon called {\it decoherence} \cite{decoherence,decoherence1,decoherence2,decoherence3}. First, any object sufficiently complex to be considered either a measuring device or observer (i.e., which, in the conventional formulation can be regarded as semi-classical) will have many degrees of freedom. Second, a measurement can only be said to have occurred if the states $M_+$ and $M_-$ differ radically: the outcome of the measurement must be stored in a redundant and macroscopically accessible way in the device (or, equivalently, in the local environment). Therefore, the ability of the second branch to affect the first branch is exponentially small: the overlap of $M_+$ with $M_-$ is typically of order $\exp( - N )$, where $N$ is a macroscopic number of degrees of freedom, and the future dynamical evolution of each branch is unlikely to alter this situation. For All Practical Purposes (FAPP, as J.S. Bell put it), an observer on one branch can ignore the existence of the other: they are said to have decohered. Each of the two observers will perceive a collapse to have occurred, although the evolution of the overall system $S'$ has continued to obey the Schrodinger equation. 

It is sometimes said that decoherence solves the measurement problem in quantum mechanics. More accurately, decoherence illuminates the process of measurement. But it does not answer the question of whether or not wave functions actually collapse -- {\it it merely makes it clear that they need not!} While decoherence is used in the most practical settings, such as the analysis of atomic experiments involving entanglement, somehow the analysts often fail to notice that their calculations work perfectly well in the absence of collapse; the measurement can proceed in a perfectly continuous manner, with the different branches rapidly losing contact with each other. The result is the appearance of a single outcome without invoking collapse.

\bigskip

The idea that wave function collapse is unnecessary, that the whole universe could be thought of as a closed system obeying the Schrodinger equation and described by a {\it universal wave function}, is due to Everett \cite{Everett,Everett1}. His conception is a {\it universal} quantum mechanics: one in which observers and measuring devices are treated on the same footing as all other quantum degrees of freedom -- there are no special objects which collapse wave functions. It is sometimes claimed that the many worlds interpretation is experimentally indistinguishable from the conventional interpretation. This is true For All Practical Purposes (FAPP \cite{Bell}), but see Deutsch \cite{Deutsch} for an experiment that is sensitive to whether or not wave functions undergo non-unitary projection. Note also that an implicit assumption of conventional quantum mechanics is that observers (i.e., objects or measuring procedures that collapse wave functions) cannot themselves been placed in superposition states. However, rapid progress is being made toward the creation of macroscopic (Schrondinger's Cat) superposition states, including, possibly, superpositions of viruses and bugs \cite{bugs}. If a bug can be placed in a superposition state, why can't you?

In fact the foundations of quantum mechanics are not entirely disconnected from practical issues in cosmology. The cosmic microwave background data favors inflationary cosmology \footnote{It is also worth noting that the while the main criticism of many worlds seems to be its {\it extravagance} (i.e., the implication that other branches exist and are equally ``real''), one could say the same about inflationary cosmology: it implies the existence of enormous (perhaps infinite) regions of space-time which are causally disconnected from us.}, according to which density perturbations in the early universe originate in the quantum fluctuations of the inflaton field itself. It is very hard to fit this into the conventional view -- what collapses the wavefunction of the inflaton? There are no observers in the early universe -- the very existence and locations of observers (such as humans) are determined by the density perturbations themselves! Galaxies, stars and planets are found in the overdense regions, but quantum mechanics itself decides which regions are overdense; there is no classical system outside or observing the universe. It seems much more natural to note that differential scattering of gravitons due to more or less energy density in a particular region separates the inflaton wavefunction into decoherent branches. The gravitons decohere the inflaton state vector through interactions. But this is accomplished through unitary evolution and does not require von Neumann projection or collapse. Other observers, living on other branches of the wavefunction, see a different CMB pattern on the sky.

\section{Probability and deterministic evolution}

The Schrodinger dynamics governing unitary evolution of the wave function is entirely deterministic. In the absence of collapse, this is the only kind of time evolution in quantum mechanics. If the initial state of $S'$ is known at some initial time $t_0$, it can be predicted with perfect accuracy at any subsequent time $t$. How, in such a deterministic system, can the notion of probability arise?

In the conventional interpretation, with only a single realized outcome of an experiment, one simply {\it imposes} the Born probability rule together with measurement collapse: the probability of the plus outcome is $|c_+|^2$, or more generally the likelihood associated with a particular component of the wave function is given by its magnitude squared under the Hilbert measure. (Philosophically, this imposition of objective randomness is a violent departure from the notion that all things should have causes; see below.) In the absence of collapse there is no logical point at which we can introduce a probability rule -- it must emerge by itself, and it must explain why an experimenter, who decoheres into different versions of himself on different branches, does so with probabilities determined by $\vert c_\pm \vert^2$. When all outcomes are realized, what is meant by the {\it probability} of a particular one? 

The problem is actually worse than this. For decoherence to work in the first place, the overlap or Hilbert space inner product between two vectors must have a probabilistic interpretation: if the inner product is small (decoherence has occurred) the two branches are assumed not to influence each other; an observer on one branch may be unaware of the other. Let us neglect this further, more subtle, complication and simply assume that decoherence and the measuring apparatus work as desired. Even in this reduced context, there is still a problem, as we now discuss.

The many worlds interpretation must provide a probability measure over all decoherent outcomes, each of which is realized, i.e., from the perspective of at least one observer or recording device. The difficulty becomes clear if we consider $N$ spins: $\Psi = \otimes_{i=1}^N ~ \psi_i$, with each spin prepared in the identical state $\psi_i = c_+ \vert + \rangle_i ~+~ c_- \vert - \rangle_i$. (See figure 2.) Again, all possibilities are realized, including the outcome where, e.g., all spins are measured in the $+$ state: $\Psi \sim \vert + + + \cdots + \rangle$. If $\vert c_+ \vert$ is small, then this outcome is very unlikely according to the usual probability rules of quantum mechanics. However, independent of the value of $c_+$, it comprises one of the $2^N$ distinct possible outcomes. Each of these outcomes implies the existence of an observer with distinct memory records. 

For $N$ sufficiently large and $|c_+| \neq |c_-|$, it can be shown \cite{BHZ} that the vast majority of the $2^N$ realized observers (i.e., weighting each distinct observer equally) see an outcome which is highly unlikely according to the usual probability rules. Note that counting of possible outcomes depends only on combinatorics and is independent of $c_\pm$. As $N \rightarrow \infty$, for all values of $c_\pm$ (excluding exactly zero), almost all of the realized observers find nearly equal number of $+$ and $-$ spins: there are many more outcomes of the form, e.g., $(+ + - + - \, \cdots \, + - - +)$ with roughly equal number of $+$'s and $-$'s than with many more of one than the other. This had to be the case, because counting of outcomes is independent of the values of $c_\pm$, leading to a symmetry between $+$ and $-$ outcomes in the combinatorics. In contrast, the Born rule predicts that the relative number of $+$ and $-$ outcomes depends on $|c_\pm|^2$. In the large $N$ limit almost all (distinct) observers \footnote{In using terms like {\it almost all} we have implicitly adopted a measure: the simple counting measure in which each of the $2^N$ observers is weighted equally.} experience outcomes that strongly disfavor the Born probability rule: {\it almost all of the physicists in the multiverse see experimental violation of the Born rule}. Or: {\it almost none of the physicists in the multiverse see outcomes consistent with the Born rule.}

Everett referred to the branches on which results deviate strongly from Born rule predictions (i.e., exhibit highly improbable results according to the usual probability rule) as {\it maverick} branches. By definition, the magnitude of these components under the Hilbert measure vanishes as $N$ becomes large. But there is no a priori sense in which the Hilbert measure is privileged in many worlds. Nor is there even a logical place to introduce it -- it must emerge in some way by itself. Everett claimed to derive quantum mechanical probability by taking $N$ to infinity and discarding all zero norm states in this limit, thereby eliminating all maverick outcomes. Most advocates of many worlds regard this reasoning as circular and look elsewhere for a justification of the Born rule.

Instead, most attempts to justify the Born rule have relied on {\it subjective} probability arguments. (Dynamical mechanisms for removing maverick branches have also been considered \cite{Hanson, BHZ}.) While objective probabilities can be defined through frequencies of outcomes of a truly random process, subjective probabilities deal with degrees of belief. The conventional quantum interpretation, with von Neumann projection, assumes true randomness and objective probabilities: repeated measurements produce a truly random sequence of outcomes, with frequencies given by the Born rule. Outcomes are unknowable, {\it even in principle}, even with perfect knowledge of the state being measured. (Einstein objected to the introduction of true randomness, because it implies outcomes without causes.) In the absence of true randomness, probabilities are subjective and represent degrees of belief \footnote{Some have characterized probability theory as a generalization of logic. See, e.g., E.T. Jaynes, {\it Probability Theory: The Logic of Science}, Cambridge University Press (June 9, 2003). The first chapter begins with the following quote: ``... Therefore the true logic for this world is the calculus of Probabilities, which takes account of the magnitude of the probability which is, or ought to be, in a reasonable man's mind." -- James Clerk Maxwell, 1850.
}.

We emphasize again that universal Schrodinger evolution only admits subjective, but not objective, randomness and probability. Just before the measurement depicted in figure 2, all of the observers are in identical states. Using the basis $\vert m \rangle \equiv \vert s_1, s_2, \cdots s_N \rangle$, where $s_i$ are individual spin eigenstates and $m$ runs from $1$ to $2^N$, we can write
\begin{eqnarray}
\label{copies}
\sum_{m}  c_m \, \vert m \rangle \otimes \vert {\cal O} \rangle ~ &=& ~ c_1 \, \vert 1 \rangle \otimes \vert {\cal O} \rangle
~+~ c_2 \, \vert 2 \rangle \otimes \vert {\cal O} \rangle ~+~ c_3 \, \vert 3 \rangle \otimes \vert {\cal O} \rangle  ~+~ \cdots   \nonumber \\
 && \longrightarrow ~~   c_1 \, \vert 1 \, , \, {\cal O}_1 \rangle
~+~ c_2 \, \vert 2 \, , \, {\cal O}_2 \rangle ~+~ c_3 \, \vert 3 \, , \, {\cal O}_3 \rangle  ~+~ \cdots ~=~
\sum_{m}  c_m \, \vert m \, , \, {\cal O}_m \rangle~~~,
\end{eqnarray}
where ${\cal O}_m$ denotes the state describing the observer who recorded outcome $m$. The first line has been written to emphasize that in the $m$ basis it appears as if there are $2^N$ identical observers, each of whom is {\it destined} to evolve into a particular one of the ${\cal O}_m$. Of course, the observer does not know which of the ${\cal O}_m$ they will evolve into, because they do not (yet) know the spin state $\vert m \rangle$ on their branch. But this is a {\it subjective} uncertainty because, indeed, the outcome is already pre-determined. This perspective may appear more natural if one considers the time reversal of the final state in (\ref{copies}). Each observer ${\cal O}_m$ evolves backward in time to one of the identical $\cal O$'s. If the ultimate decoherent basis of the universe were known a priori, the evolution of each branch would appear entirely deterministic \footnote{The {\it reason} that observer ${\cal O}_m$ observed outcome $m$ is that, before the measurement, he was paired with the spin state $\vert m \rangle$: $~\, \vert m \rangle \otimes \vert {\cal O} \rangle ~\rightarrow~ \vert m \, , \, {\cal O}_m \rangle$. This initial condition {\it caused} the measurement outcome. Compare to the conventional formulation, in which, e.g., the observer measuring spin state $( \vert + \rangle \, + \, \vert - \rangle )$ finds the outcome $+$ and the wave function collapses to $\vert + \rangle$ only, with no $\vert - \rangle$ branch. Clearly there is no {\it cause} for the $+$ outcome (as opposed to $-$): the result is truly, objectively random, something that Einstein objected violently to.} \cite{Csonka}. Of course, this still neglects the question of why {\it my consciousness} in particular has been assigned to a specific decoherent branch of the universal wave function. But this question, which singles out a particular consciousness out of the many that are presumed to be equally real, is in a sense beyond the many worlds framework.

\bigskip

Subjective probability arguments originate in certain postulates governing the way in which we {\it reason} about probabilities \cite{zurek}. For example, following Laplace, we might require that two components of the wave function related by symmetry (i.e., with equal coefficients $c_i$ and $c_j$) must have equal probabilities as outcomes. By further analysis, we might conclude that probability must be proportional to Hilbert space norm squared. Indeed, Gleason's theorem \cite{gleason} suggests that any reasonable calculus of quantum probabilities must result in the familiar Born rule. The reader can judge for himself whether these arguments are convincing \footnote{Personally, I find the arguments convincing as a prescription for how a ``rational'' person should reason probabilistically within the framework of many worlds quantum mechanics. However, not every rational person would be able to reproduce the results! (See L. Vaidman's {\it Quantum Sleeping Beauty} thought experiment for an example where experts disagree as to probability calculus: http://philsci-archive.pitt.edu/8558/ .) I am still discomfited by the numerical (i.e., according to the counting measure) predominance of maverick branches. But perhaps I shouldn't be :-)}. It is important to emphasize that, even if successful, such arguments only address whether a particular observer should be {\it surprised} to see the Born rule at work on his branch of the wave function. They do not alter the fact that the many worlds wave function realizes almost exclusively observers who see gigantic violations of the Born rule (even, gross departures from decoherence). This vast majority of observers seem unlikely to believe in quantum mechanics as the correct theory of Nature, let alone to conform to the reasoning described above. The subjective arguments reassure us that we are ``unlikely'' to be one of these observers. But are we reassured?

\begin{figure}
\includegraphics[width=16cm]{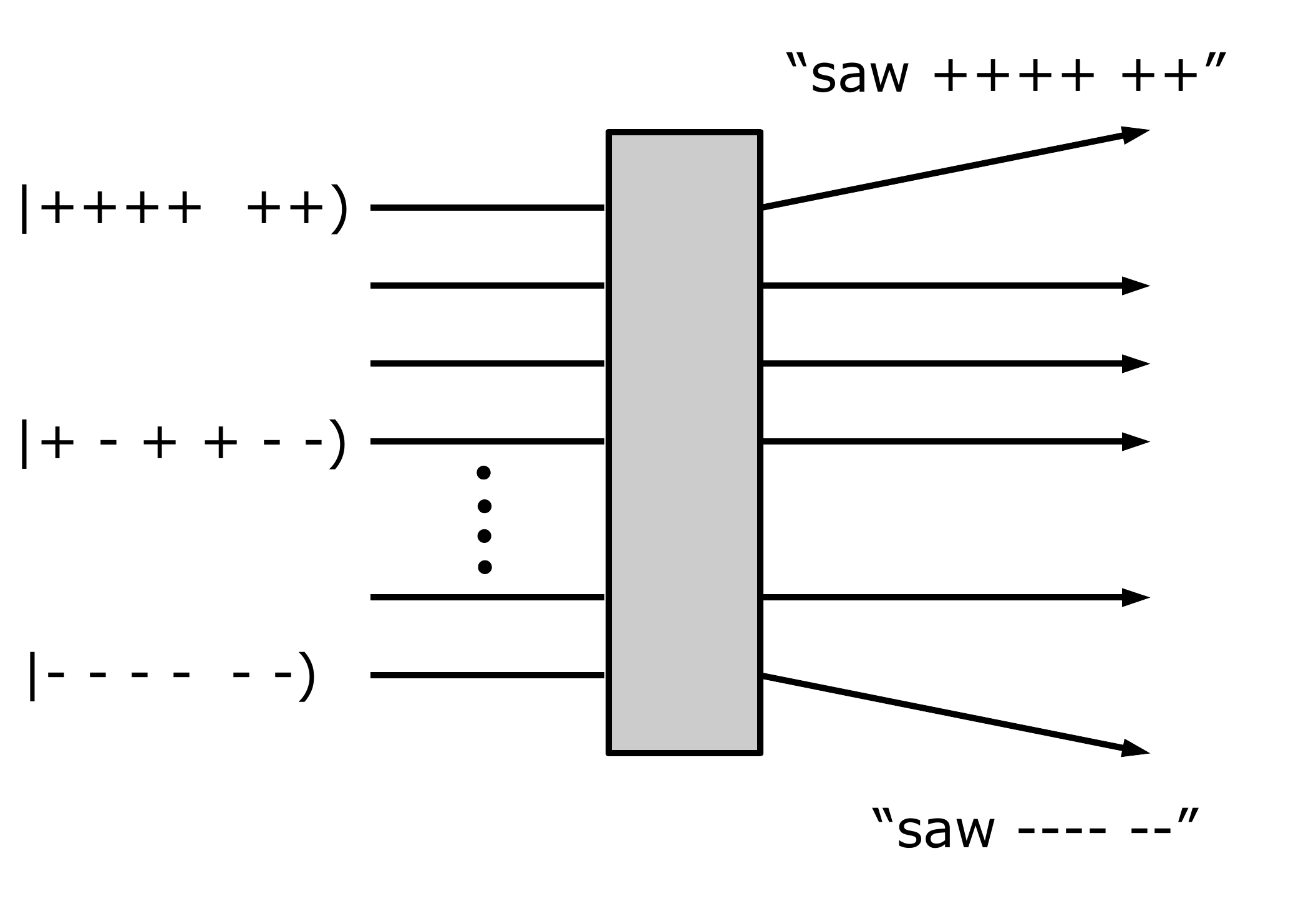}
\caption{An observer measures a state comprised of $N$ spins. Before the measurement there is a single observer state, but afterwards there are $2^N$ observer states.}
\label{identity2}
\end{figure}

\bigskip

\section{Conclusions}

Decoherence does not resolve the collapse question, contrary to what many physicists think. Rather, it illuminates the process of measurement and reveals that pure Schrodinger evolution (without collapse) can produce the quantum phenomena we observe. This of course raises the question: do we need collapse? If the conventional interpretation was always ill-defined (again, see Bell for an honest appraisal \cite{Bell}; Everett referred to it as a ``philosophical monstrosity''), why not remove the collapse or von Neumann projection postulates entirely from quantum mechanics?

The origin of probability is the real difficulty within many worlds interpretations. The problem is subtle and experts are divided as to whether it has been resolved satisfactorily. Because the wave function evolves entirely deterministically in many worlds, all probabilities are necessarily subjective and the interpretation does not require true randomness, thereby preserving Einstein's requirement that outcomes have causes.

\bigskip
\bigskip
\bigskip

\emph{Acknowledgements---} The author thanks S. Lloyd, L. Maccone, D. Politzer, J. Preskill, D. Reeb, A. Zee, D. Zeh, and especially W. Zurek for useful discussions. The author is
supported by the Department of Energy under DE-FG02-96ER40969. This essay is based on several talks given over the years, including at the Benasque Center for Science (workshop on quantum coherence and decoherence),  Caltech Institute for Quantum Information, National Taiwan University and Academia Sinica, Taiwan.

%%%%%%%%%%%%%%%%%%%%%%%%%%%%%%%%%%%%%%%%%%%%%%%%%%%%%%%%%%%%%%%%%
%%%
%%%                     BIBLIOGRAPHY
%%%
%%%%%%%%%%%%%%%%%%%%%%%%%%%%%%%%%%%%%%%%%%%%%%%%%%%%%%%%%%%%%%%%%

\bigskip

%\newpage
%\vskip .75 in

\end{document}